\documentclass[journal]{IEEEtran}

\usepackage{graphicx}
\usepackage{balance}
\usepackage{comment}
\usepackage{fixltx2e}
\usepackage{amsmath}
\usepackage{balance}
\usepackage{comment}
\usepackage{subfigure}
\usepackage{epsfig}
\usepackage{url}
\usepackage{amsfonts}
\usepackage{amssymb}
\usepackage{mdwmath}
\usepackage{mdwtab}
\usepackage{cite}
\usepackage{booktabs}
\usepackage{tabularx}
\usepackage{multirow} 
\usepackage[]{caption}
\usepackage{amsmath}
\usepackage{hyperref}
\usepackage[fleqn]{mathtools}
\usepackage{graphicx}
\usepackage{relsize}
\usepackage{flushend}
\usepackage{booktabs}
\usepackage{adjustbox}

\begin{document}

\title{\huge{Optimization for Infrastructure Cyber-Physical Systems}}

\author{Arunchandar Vasan,
        Prasant Misra,
        Srinarayana Nagarathinam,
        Venkata Ramakrishna,
        Ramasubramanian Suriyanarayanan,        
        and~Yashovardhan Chati 
\thanks{The authors are with Tata Consultancy Services - Research, India.}
\thanks{}
\thanks{}}

\maketitle

%
\IEEEpeerreviewmaketitle

\section{Introduction}

Cyber-physical systems (CPS) are systems where a decision making (cyber/control) component is tightly integrated with a physical system (with sensing/actuation) to enable real-time monitoring and control. Recently, there has been significant research effort in viewing and optimizing physical infrastructure in built environments as CPS, even if the control action is not in real-time. Some examples of infrastructure CPS include electrical power grids; water distribution networks; transportation and logistics networks; heating, ventilation, and air conditioning (HVAC) in buildings; etc. Complexity arises in infrastructure CPS from the large scale of operations; heterogeneity of system components; dynamic and uncertain operating conditions; and goal-driven decision making and control with time-bounded task completion guarantees. For control optimization, an infrastructure CPS~\cite{annaswamy2016:cps,misra2013:cps} is typically viewed as a system of semi-autonomous sub-systems with a network of sensors and uses distributed control optimization to achieve system-wide objectives that are typically measured and quantified by better, cheaper, or faster system performance. In this article, we first illustrate the scope for control optimization in common infrastructure CPS. Next, we present a brief overview of current optimization techniques. Finally, we share our research position with a description of specific optimization approaches and their challenges for infrastructure CPS of the future.

\section{Operations Optimization for Infrastructure Systems}

\subsection{Electricity Grids}

Electrical grids are complex systems that involve control optimization at multiple levels~\cite{taneja2012:grid-cps} including the following:  
\vspace{1mm}
\newline
\noindent
$\bullet$ \textbf{Grid level control}: The main optimization challenge here is to ensure that a heterogeneous mix of generation power sources matches a time-varying customer demand for electricity using predicted demand (e.g., from analytics using weather data) and the feedback signal of power quality (sensed at demand points through PMUs) to obtain real-time actuation of a heterogeneous mix of generation power sources, even while ensuring grid stability and availability. Because power generators are physical systems with significant inertia, they have start-up delays and are prone to myriad thermomechanical faults and failures. Additionally, the control decisions are constrained by the transmission and distribution limitations of the grid. 
\vspace{1mm}
\newline
\noindent
$\bullet$ \textbf{Supply side control}: The key optimization challenges within the generation sources at the supply side are typically to ensure efficient operations and improve the asset lifetime. As an example, wind turbines~\cite{venkat2022:falcon} need continuous (typically, pitch and yaw) control to ensure maximum power yield from rapidly changing wind conditions and a longer life of assets (e.g., gearbox, blade health). Further, a wind farm-level controller could potentially make better decisions than an individual turbine level.  In solar plants with thousands of photovoltaic cells, trackers could follow the movement of the sun to extract the maximum solar irradiance through the day. 
\vspace{1mm}
\newline
\noindent
$\bullet$ \textbf{From supply to demand}: Because the prime movers (wind, sun, tides, etc.,) in renewable sources are highly unpredictable, integrating renewables in an electricity grid requires the grid to handle additional variance at the supply side in addition to the existing variance at the demand side. Therefore, if renewables are to constitute a significant fraction of supply, alternate grid models are required. Specifically, instead of the current supply-following-demand model, optimization approaches are needed to orchestrate/control a demand-following-supply model where demand is shifted from periods of energy deficit to those of higher supply. 
\vspace{1mm}
\newline
\noindent
$\bullet$ \textbf{Demand side control}:  The demand side in the electricity grid is dominated by large buildings. In buildings, HVACs that account for $50$\% of the electricity consumption are main targets for control optimization. The key control challenge here is to primarily ensure user comfort and biosafety in an energy and cost-efficient manner; and secondarily, participate in demand-side management control (e.g., through demand-response programs) at the grid level to support the newer demand-following-supply model~\cite{manoharan2021:hvac-cps}. Because building demand is driven by human activity, privacy preserving occupancy detection is a key sensing challenge. Further, the control optimization needs to handle both static heterogeneity and dynamic age-related degradation in chiller capacities. Any model-predictive approach for building control requires development and (re)calibration of building thermal models that can be challenging due to ambient infiltration, solar radiation, and surface temperatures that are typically hard to measure. Finally, each building is unique in its static design and dynamic demand characteristics (much like its human occupants). 

\subsection{Water Distribution Networks}

A water supply system consists of infrastructure that collects, treats, stores, and distributes water between water sources and consumers. The distribution network manages the delivery operations where water is routed to the end consumers at the appropriate quality, quantity, and pressure via a complex network of pipes. Non-revenue water loss is a prime problem in the delivery network. This is water that does not make it from the source to the destination due to leakage, breaks or theft. It is difficult to locate the specific leak points in the pipe network and identify the cause since pipes are typically laid below the surface where it cannot be easily reached. Optimization challenges here include dynamic leak detection and management from limited sensing (sparsely located pressure/flow meters) and actuation capabilities (pressure/pumping control)~\cite{narayanan2014:water-cps}. 

\subsection{Transportation Networks}

Transportation systems no longer just deliver point-to-point travel, but have evolved into a more complex, dynamic mobility ecosystem that delivers on-demand service. Specifically, this new mobility model requires control, optimization, and orchestration across multiple service providers such as transportation, parking, electric vehicle charging and refueling, payment, remote assistance, media, food and utilities, regulations, etc. The key challenge here is to provide a seamless user experience journey across multi-modal transportation platforms through dynamic optimization~\cite{misra2020:cps,narendra2016:mobileiot}. 
\newline
\indent
Further, increasing user demand for transportation that is more uncertain in both space and time, cannot be handled by scaling the city’s transportation supply due to shortage of undeveloped urban space and significant costs for new public infrastructure. This demand could be better managed through resource uberization. Specifically, resource uberization can augment existing public capacity with privately held resources (e.g., make a private parking lot or charging point available for public consumption). In this case, the supply side also has spatiotemporal variation that need not be in sync with spatiotemporal variations in demand for mobility. This mismatch in the supply and demand is an opportunity for suitable control optimization. 

\section{Classical Optimization Methods for Infrastructure CPS}

For optimizing the operations of a physical process, a good model of the process is needed. A model focuses on the important elements of the process and abstracts out the rest. Models can be mathematical (i.e., based on mathematical functions derived either analytically or empirically) or simulation based. For infrastructure CPS, the output of the model’s optimization (‘decision variables’) would be the controller’s inputs to achieve a desired feature (‘objective function’) of the controlled process. Appropriate numerical or analytical optimization methods are used to solve the optimization problem where the respective method is used to choose the best solution among various available alternatives, while keeping in line with the objective function of the system process and the operating constraints. Depending on the forms of the objective function and constraints, optimization problems can be classified as linear, quadratic, semi-definite, semi-infinite, integer, non-linear, etc. Classical methods are broadly of the following types~\cite{sun2020:opt}:
\begin{itemize}
 \item \emph{First order}: They are iterative techniques that only use the information provided in the objective function and its gradient (or sub-gradient) to find the optimal solution. Examples include the (batch and stochastic) gradient descent, AdaGrad, Adam, etc.
 \item \emph{Higher order}: They use curvature information, in addition to the gradient, to solve problems where the objective function is highly non-linear and ill-conditioned. Examples include conjugate gradient, (stochastic) quasi-Newton method, Hessian-Free method, etc.
 \item \emph{Derivative-free}: They are used in problem settings where the derivative of the objective function and constraints may not exist or is difficult to calculate. Examples include heuristic (that are characterized by empirical rules); metaheuristic (such as genetic algorithms, simulated annealing, differential evolution, harmony search, ant colony algorithms, and particle swarm optimization).
\end{itemize}
Classical optimization approaches have several challenges. First, models need to be calibrated and validated. Calibrating physical models is well-known to be difficult and data intensive. Second, even well-calibrated models may not work well with time due to degradation of infrastructure systems. Third, a larger system model composed of several calibrated sub-system models may not be representative of the larger system unless carefully validated. This is because the modeling errors of the sub-systems while individually sufficient may compound in unforeseen ways. Finally, classical optimization methods may not be computationally scalable for deciding control actions within an available decision window. 

\section{New Optimization Approaches for Infrastructure CPS}

Key desiderata of future optimization approaches for infrastructure CPS include \emph{intelligence, composability, scalability, and practicality}. 
\newline
\indent
In addition to standard control of a CPS, intelligence in a CPS will be characterized by an ability to learn and improve at tasks over time, either from an individual interaction with the environment or from collective experiences through centralized over-the-air (OTA) knowledge sharing. For example, an autonomous vehicle (AV) can learn to drive and share its experiences with a centralized database, where post processing, an OTA update to the car’s driving ability occurs. A CPS typically involves sequential decision making, and so reinforcement learning is a popular candidate for learning to control a CPS. Because a CPS could be mission critical, learning typically may not be on a real-world system. Instead, a digital twin is typically used in a simulated environment for sufficient exploration and such learning is transferred to the real-world with limited experimentation. Edge computing and machine learning will likely be key mechanisms for delivering such learning technologies in typical CPS. Unlike purely software systems, CPS needs to respect physical constraints such as inertia and lag in both sensing and actuation. For instance, a wind turbine under rotation may not be able to change its pitch and yaw at a rate higher than some upper bound. 
\newline
\indent
A CPS can become truly useful when it can be composed with or be part of a larger system that involves identical or other systems. For identical CPS, self-organization is a key requirement that can help build scale on demand, for example with an aerial drone survey of all or part of an entire city. Swarm intelligence from nature could be a guiding principle for such architectures. A smart city is a good example of a large-scale system composed of multiple non-identical smaller CPS that together deliver value to citizens. 
\newline
\indent
Because the whole is typically much more than the sum of its parts, scalability is needed specifically in decision-making involving multiple CPS systems. Specifically, computation for decision-making problems that may be optimally solvable at the level of one CPS (e.g., how to charge one electric vehicle (EV) or control one wind turbine), may not be tractable when considering an overall composed system (e.g., how to control a farm of turbines optimally or charge a fleet of EVs for delivery). Such problems would typically require a mix of learning and classical optimization. Specifically, reinforcement learning at a higher level of abstraction (e.g., schedule this task on the CPS first) with constrained optimization (e.g., ensure that resource bounds are not violated) could help break the curse of dimensionality. 
\newline
\indent
Finally, for CPS research to have real-world impact, it needs to respect the boundaries of original equipment manufacturer (OEM) implementations, even while pushing the boundaries of innovation. For example, a learned control strategy for HVACs would ideally override the typical OEM implementation of PID type control for fan-speed in building management systems. However, for OEMs to implement the ideas in products, the learned control would set the targets for the PID controllers to achieve – as a sort of meta-controller. Such meta-control would inevitably give way to learned control gradually as OEMs see the benefits over time.  

\section{Final Remarks}
Cyber-physical systems are now becoming increasingly prevalent, and their application in built environments and urban infrastructure is becoming mainstream. In this article, we presented the opportunities for system optimization in such settings. We discussed various classical optimization techniques and newer approaches that can be applied to solve operational problems in the infrastructure CPS domain.

\balance
\begin{footnotesize}
\bibliographystyle{IEEEtran}
\bibliography{newsletter_references}
\end{footnotesize}

\balance
\vspace{-5mm}
\begin{IEEEbiography}[{\includegraphics[width=1in,height=1.25in,clip,keepaspectratio]{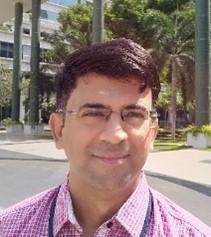}}]{Arunchandar Vasan} is a Principal Scientist at Tata Consultancy Services - Research. He obtained his BTech degree in Computer Science and Engineering from IIT Madras, India, and his MS degree and PhD in Computer Science from the University of Maryland, USA. His current research interests are in intelligent physical infrastructure enabled through cyber-physical systems.
\end{IEEEbiography}
\vspace{-5mm}
\begin{IEEEbiography}[{\includegraphics[width=1in,height=1.25in,clip,keepaspectratio]{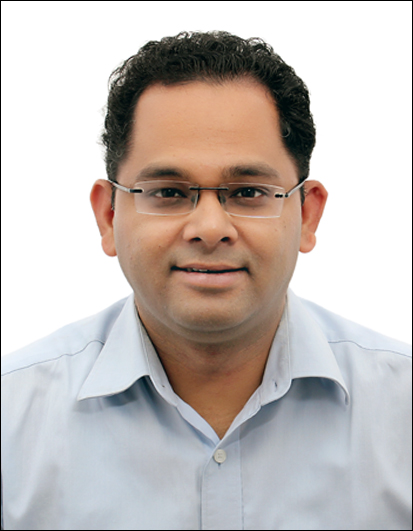}}]{Prasant Misra} is a Senior Scientist at Tata Consultancy Services - Research, where he works on intelligent cyber-physical systems for electric mobility. He received his PhD in Computer Science and Engineering from UNSW Sydney and completed his Post-doctoral from SICS Stockholm. He has received several recognitions for his work, of which it is noteworthy to mention MIT TR35 India (2017), the ERCIM Alain Bensoussan and Marie Curie Fellowship (2012), and the Australian Government's AusAID Australian Leadership Awards (2008). He is the Chair of IEEE Computer Society Bangalore Chapter; Vice-Chair (Industry Engagement) of IEEE Bangalore Section; member of the executive committee of COMSNETS Association. He is a senior member of the IEEE and ACM.
\end{IEEEbiography}
\vspace{-5mm}
\begin{IEEEbiography}[{\includegraphics[width=1in,height=1.25in,clip,keepaspectratio]{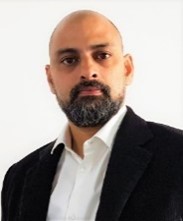}}]{Srinarayana Nagarathinam} is a Principal Scientist at Tata Consultancy Services - Research, where he leads the biosafe and energy-smart building program. He holds a PhD from the Department of Mechanical Engineering, University of Sydney. He is a recipient of the prestigious postdoctoral fellowship award by the Department of Science and Technology of the Australian government. He has extensive experience in thermal management of building systems, computational fluid dynamics, and heat transfer. His recent research interests include the application of physics-based and ML/AI-based techniques to optimal control of HVAC systems.
\end{IEEEbiography}
\vspace{-5mm}
\begin{IEEEbiography}[{\includegraphics[width=1in,height=1.25in,clip,keepaspectratio]{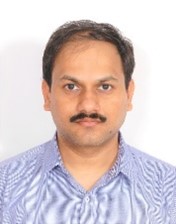}}]{Venkata Ramakrishna} obtained his PhD in Mechanical Engineering from IIT Madras. He is currently a Senior Scientist at Tata Consultancy Services - Research with 15 years of research experience in the areas of Renewable energy, Sustainability, Electric grids, Energy systems Modeling \& Analytics. His current research involves developing Digital Twins \& Solutions for Wind and Solar power plants. He leads multi-disciplinary research team which develops domain models, uses simulations, data mining, AI and CS algorithms. He published 32 papers and filed 8 patents with 4 patents granted. He serves as reviewer for Elsevier, IEEE, ASME, ACM and Hindawi publications. He is an academic editor for Int. J. Electrical Components and Energy.
\end{IEEEbiography}
\vspace{-5mm}
\begin{IEEEbiography}[{\includegraphics[width=1in,height=1.25in,clip,keepaspectratio]{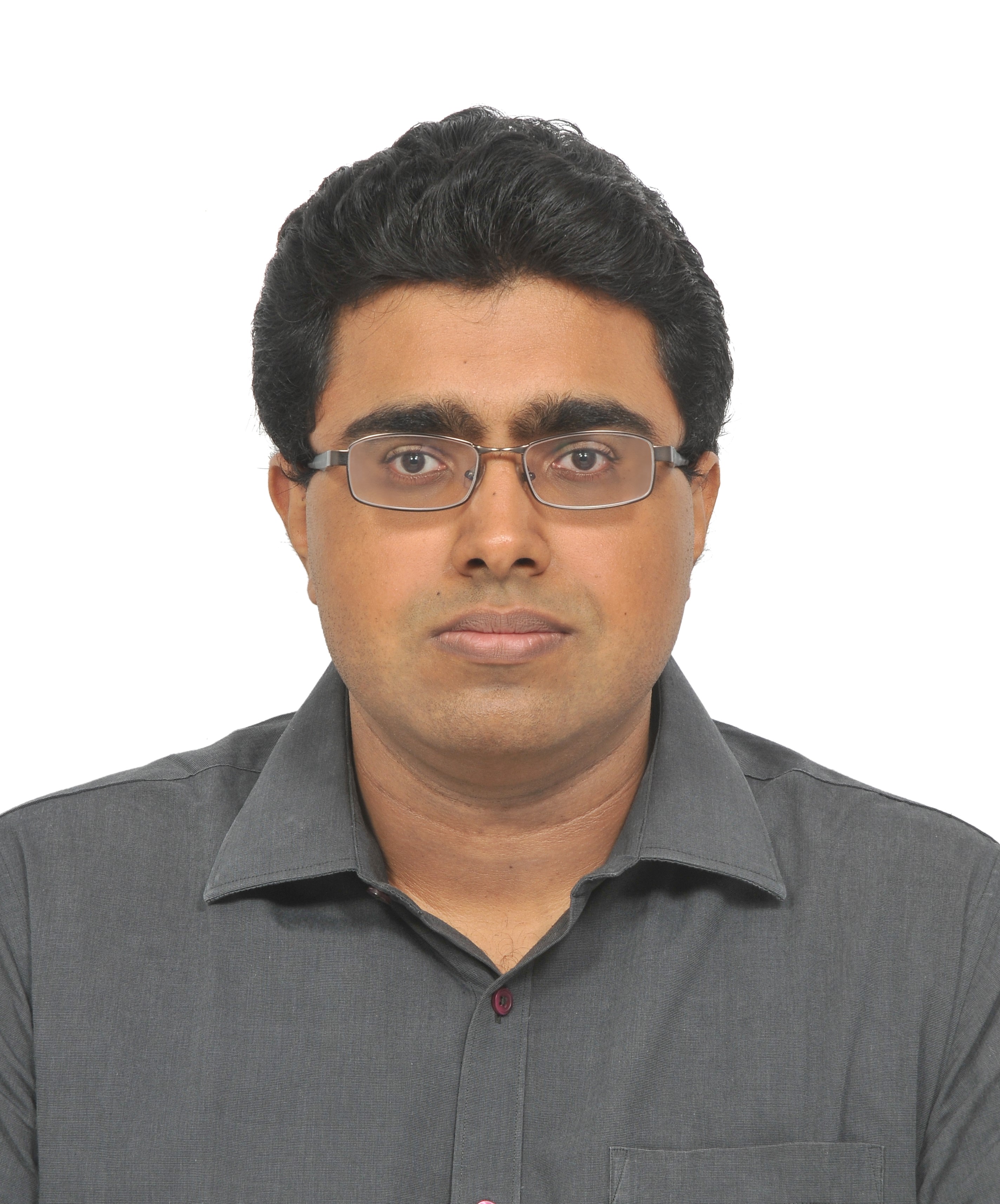}}]{Ramasubramanian Suriyanarayanan} is a Solution Architect at Tata Consultancy Services - Research. He obtained his BTech degree in Computer Science and Engineering from Sastra University, India. He has 14 years of research experience in the areas of Energy, Sustainability, Compliance and Aviation. His current research interests are towards optimizing airline and airport operations using ML/AI-based techniques and simulation.
\end{IEEEbiography}
\vspace{-5mm}
\begin{IEEEbiography}[{\includegraphics[width=1in,height=1.25in,clip,keepaspectratio]{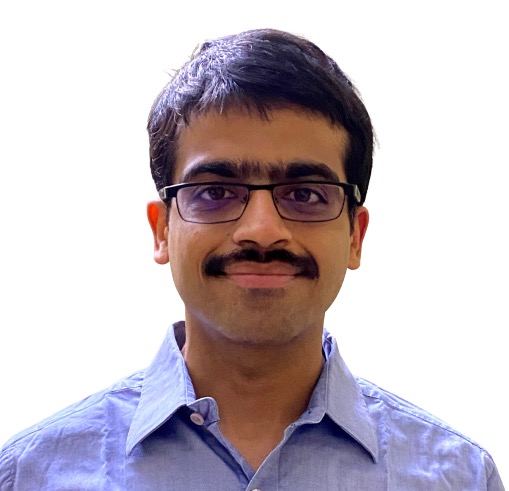}}]{Yashovardhan Chati} is a Scientist at Tata Consultancy Services - Research., where he broadly works on developing physics and data-driven models to analyze and control diverse cyber-physical systems. He obtained his BTech and MTech degrees in Aerospace Engineering from IIT Bombay. He pursued his PhD in Aeronautics and Astronautics at the Massachusetts Institute of Technology, USA where he developed machine learning algorithms to model aircraft performance. He is a recipient of the IIT Bombay Institute Gold Medal in 2012, the US Transportation Research Board’s ACRP Graduate Research Award on Public Sector Aviation issues in 2016, and a best paper-in-track award at the USA/Europe Air Traffic Management Research and Development Seminar 2017.
\end{IEEEbiography}

\end{document}